\documentclass[aps,prb,reprint,superscriptaddress,amsfonts,amssymb,amsmath,10pt]{revtex4-1}

\usepackage{graphicx}
\usepackage{hyperref}
\usepackage{mathtools}

\usepackage{dcolumn}
\usepackage{bm}
\usepackage{xfrac}
\usepackage{SIunits}
\usepackage{color}
\usepackage[usenames,dvipsnames]{xcolor}
\usepackage{footmisc}

\usepackage{subfigure}

\begin{document}

\title{A spin dynamics study in layered van der Waals single crystal, Cr$_2$Ge$_2$Te$_6$}

\author{S. Khan}
\email{safe.khan.11@ucl.ac.uk}
\affiliation{London Centre for Nanotechnology, University College London, London WC1H 0AH, United Kingdom}

\author{C. W. Zollitsch}
\affiliation{London Centre for Nanotechnology, University College London, London WC1H 0AH, United Kingdom}

\author{D. M. Arroo}
\affiliation{London Centre for Nanotechnology, University College London, London WC1H 0AH, United Kingdom}

\author{H. Cheng}
\affiliation{Department of Physics, University of Science and Technology Beijing, Beijing 100083, China}

\author{I. Verzhbitskiy}
\affiliation{Department of Physics, National University of Singapore, 2 Science Drive 3, Singapore 117551}

\author{A. Sud}
\affiliation{London Centre for Nanotechnology, University College London, London WC1H 0AH, United Kingdom}

\author{Y.P. Feng}
\affiliation{Department of Physics, National University of Singapore, 2 Science Drive 3, Singapore 117551}

\author{G. Eda}
\affiliation{Department of Physics, National University of Singapore, 2 Science Drive 3, Singapore 117551}

\author{H. Kurebayashi}
\email{h.kurebayashi@ucl.ac.uk}
\affiliation{London Centre for Nanotechnology, University College London, London WC1H 0AH, United Kingdom}

\begin{abstract}

We study the magnetisation dynamics of a bulk single crystal Cr$_2$Ge$_2$Te$_6$ (CGT), by means of broadband ferromagnetic resonance (FMR), for temperatures from 60 K down to 2 K. We determine the Kittel relations of the fundamental FMR mode as a function of frequency and static magnetic field for the magnetocrystalline easy - and hard - axis. The uniaxial magnetocrystalline anisotropy constant is extracted and compared with the saturation magnetisation, when normalised with their low temperature values. The ratios show a clear temperature dependence when plotted in the logarithmic scale, which departs from the predicted Callen-Callen power law fit of a straight line, where the scaling exponent \textit{n}, $K_{u}(T) \propto [M_s(T)/M_s(2$ K$)]^n$, contradicts the expected value of 3 for uniaxial anisotropy. Additionally, the spectroscopic g-factor for both the magnetic easy - and hard - axis exhibits a temperature dependence, with an inversion between 20 K and 30 K, suggesting an influence by orbital angular momentum. Finally, we qualitatively discuss the observation of multi-domain resonance phenomena in the FMR spectras, at magnetic fields below the saturation magnetisation. 

\end{abstract}
\date{\today}
\maketitle

\section{Introduction}

Two-dimensional (2D) van der Waals (vdWs) single crystals, belonging to the family of lamellar ternary  chalcogenides (i.e$.$ CGT and Cr$_2$Si$_2$Te$_6$) and chromium halides (i.e$.$ CrI$_3$ and CrBr$_3$), have recently attracted a great deal of interest due to the presence of long range magnetic order in the 2D limit\cite{liu2016critical, wang2018very,richter2018temperature,gong2017discovery}. The presence of ferromagnetism in the 2D state has a potential to open new avenues in the field of spintronics leading to new magneto-optical and magneto-electric applications\cite{cardoso2018van,fang2018large}. Recent study of gate-tunable room temperature ferromagnetism in layered 2D Fe$_3$GeTe$_2$ \cite{deng2018gate}and the discovery of near room temperature ferromagnetism in the cleavable Fe$_5$GeTe$_2$ \cite{may2019ferromagnetism} have highlighted the  significance that these layered vdWs systems can have for spintronics devices with room temperature applications. Thermal fluctuations in 2D systems at finite temperatures can restrain the long range magnetic order according to the Mermin-Wagner theorem\cite{mermin1966nd}, however due to the presence of large magnetocrystalline anisotropy in these layered systems the magnetic order remains dominant down to a few layers.

Chromium tellurogermanate, CGT, is a layered 2D ferromagnetic semiconductor with vdWs coupling between the adjacent layers. Bulk CGT has a hexagonal crystal structure with the R$\bar{3}$ space group \cite{lin2017effects}. CGT has been a subject of vast experimental studies over the past few years. Gong et al. discovered the intrinsic ferromagnetism in atomic bilayers of CGT using the magneto-optical technique and showed a significant control between paramagnetic to ferromagnetic transition temperature with very small magnetic fields\cite{gong2017discovery}. For CGT with only few layers ($\approx 3.5$ nm thickness of crystalline flakes), Wang et al. demonstrated the control of magnetism by an electric field, showing a possibility for new applications in 2D vdWs magnets\cite{wang2018electric}. Liu et al. report an anisotropic magnetocaloric effect associated with the critical behaviour of CGT and provide evidence of 2D Ising-like ferromagnetism, which is preserved in few-layer devices\cite{liu2019anisotropic}. There still remains an ambiguity in the type of magnetic interaction in CGT, as experimentally, Heisenberg-like ferromagnetism is reported\cite{gong2017discovery}, but other reports looking at the critical exponents in CGT predict 2D Ising-like ferromagnetism\cite{liu2017critical, liu2018critical}.

CGT has been proposed as a potential substrate for topological insulators in order to realise the quantum anomalous hall effect, and a large anomalous hall effect in the bilayer structure of Bi$_2$Te$_3$ and CGT has been already observed\cite{alegria2014large}. To this point, there has been only one brief report of determination of the uniaxial magnetic anisotropy in CGT by ferromagnetic resonance (FMR)\cite{zhang2016magnetic}, while recent experiments rely on probing the sample by means of magneto-optical, magnetometry and transport techniques. Understanding the magnetisation dynamics by FMR is beneficial as it exactly measures the magnetic ground state\cite{anisimov1999orbital} i.e. uniform-mode excitations of spin-waves with $k \approx 0$. The FMR experiment allows to determine the magnetic anisotropies in the system, the spectroscopic splitting g-factor, and addtionally it can provide information on the relative orbital contribution to the magnetic moment\cite{farle1998ferromagnetic}. Therefore, it is significant to better understand the magnetisation dynamics in bulk CGT in order to unfold the full potential of these layered vdWs systems in the field of spintronics.

In this article, we report on a broadband FMR study in CGT in the temperature range of 60 K - 2 K, with the external magnetic field applied along the in-plane (ab-plane) and out-of-plane (c-axis) orientations. The extracted value of the uniaxial magnetocrystalline anisotropy constant, $K_{\textrm{u}}$, is found to be temperature-dependent. We find that the scaling of magnetic anisotropy constant and saturation magnetisation as a function of temperature deviates from the theoretical prediction by the Callen-Callen power law. The determined g-factor in CGT is found to be anisotropic for the different crystallographic directions. Finally, we observe a domain-mode resonance phenomenon below the saturation field. This indicates a presence of multiple domain structures in CGT.

\section{Experimental Details}

CGT single-crystalline flakes were synthesized by the direct vapour transport (or flux) method. High-purity elemental Cr (99.9999\% in chips), Ge (99.9999\% in crystals) and Te (99.9999\% in beads) were pre-mixed at molar ratio of 20:27:153 and sealed in the quartz ampule at high vacuum ($\approx 10^{-5}$ Torr). The ampule was loaded in the single-zone furnace, heated up to 1273 K with the rate of 2 K/min and left for 2 days. For a low-defect crystal growth, the furnace was set to slow cooling with the rate of 5 K/hour down to 673 K. Then, the furnace was turned off for natural cooling. The layered single-crystalline flakes were separated from the flux build-up and stored in the inert atmosphere of the glovebox. The bulk single crystals of CGT are not air sensitive and it is also the case for our sample whereas it is different for few layers of the CGT where the bilayer degrades after 90 mins of exposure to air \cite{gong2017discovery}.

Magnetisation measurements were carried out at 2K and 60K with a vibrating sample magnetometer in a Quantum Design PPMS-14T, with measurements taken for both in-plane and out-of-plane applied magnetic fields. The Curie temperature of CGT is determined through heat capacity measurements carried out in the absence of any magnetic fields by thermal relaxation calorimetry in the PPMS\cite{lashley2003critical}.

A standard broadband FMR technique is used for the study of magnetisation dynamics in CGT \cite{he2016broadband,bilzer2007vector,montoya2014broadband}. To employ a broadband oscillatory magnetic field to the sample, a silver plated copper co-planar waveguide (CPW) on a Rogers 4003c PCB, with a center conductor width of 1 mm and a gap of 0.5 mm, is used. The impedance matching of the microwave circuitary connnected to the CPW is important for consideration when using it for FMR experiments, and it is carefully matched to 50 $\Omega$ \cite{pozar2009microwave}. The PCB is mounted inside a cooper sample box and features two SMP connectors, allowing the connection to the microwave circuitry. The sample box is attached to a probe, which is fitted inside a closed cycle helium flow cryostat, with a temperature range of 2 K - 300 K. In addition, the cryostat is equipped with a rotation stage, allowing static magnetic field angle dependence measurements. The static magnetic field is generated by an electromagnet (Bruker ER073) which is able to produce a highly homogeneous field at the sample location.

We perform broadband microwave transmission experiments, using a vector network analyzer (VNA, Hewlett Packard). The microwave transmission S$_{21}$ is measured as a function of VNA frequency and static magnetic field. The FMR signature is identified by a strong microwave absorption, hence a reduction of transmission.

\section{Results \& Discussion}

\begin{figure}[htb!]
	\centering
	\includegraphics[width=1.0\linewidth]{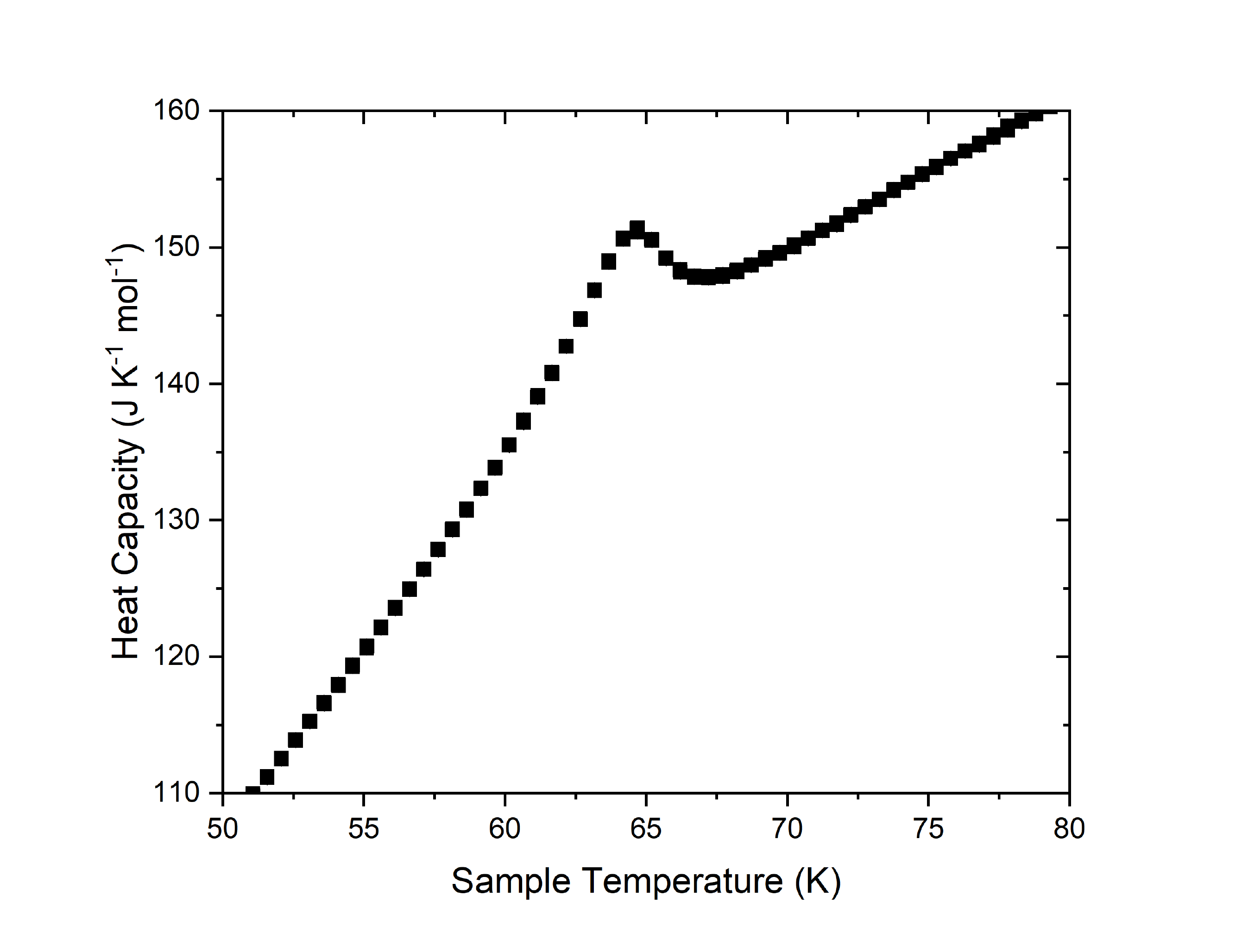}
	\caption{Heat capacity of a CGT sample measured by thermal relaxation calorimetry. The anomaly in the heat capacity associated with the paramagnetic to ferromagnetic transition indicates a Curie temperature of 64.7 $\pm$ 0.5 K.}
	\label{fig:heatcap-cgt-50k-80k}
\end{figure}

The Curie temperature of the CGT sample is determined to be $T_{\textrm{c}} = 64.7\pm 0.5$ K via a characteristic peak in the heat capacity associated with magnetic ordering (Fig. \ref{fig:heatcap-cgt-50k-80k}). The cusp in the heat capacity data as a function of sample temperature corresponds to collinear magnetic ordering in CGT, as reported elsewhere \cite{liu2018critical}.

\begin{figure*}
	\centering
	\includegraphics[width=1.0\linewidth]{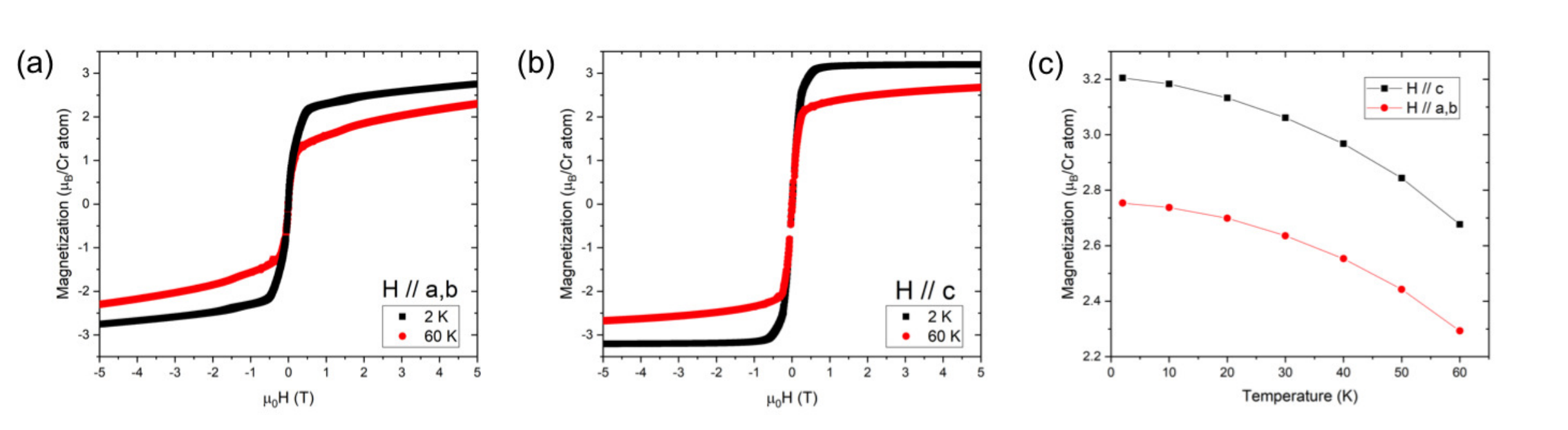}
	\caption{Magnetisation data for CGT measured via vibrating sample magnetometry carried out at temperatures from 2 K to 60 K. (a) Magnetisation curves with the field applied along the in-plane direction. (b) Magnetisation curves with the field applied along the out-of-plane direction. (c) Magnetisation with an applied field of 5 T as a function of temperature, lines connecting data points have been added as a guide to the eye.}
	\label{fig:ppms-vsm-cgt-all}
\end{figure*}

The magnetisation of the CGT sample is measured below the Curie temperature for fields along the in-plane and out-of-plane directions, yielding respective saturation magnetisations of 2.8 and 3.2 $\mu_{\textrm{B}}$ per Cr atom at 2 K (Fig. \ref{fig:ppms-vsm-cgt-all}).

\begin{figure}[htb!]
\centering
\includegraphics[width=1.0\linewidth]{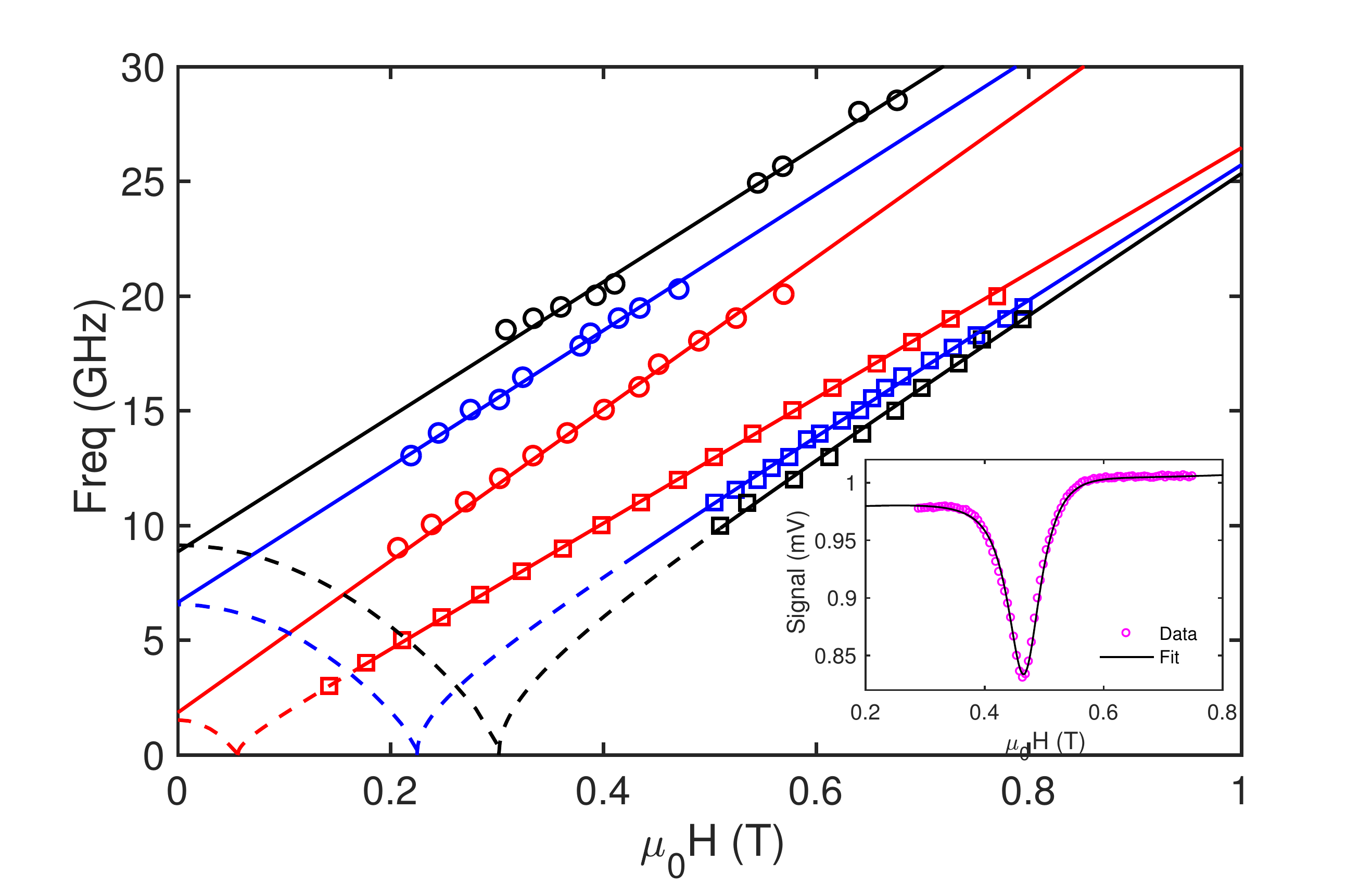}
\caption{The inset shows the experimental resonance spectra with a Lorentzian line shape (black line). Main plot shows frequency dependence of resonance field at 60 K (red), 30K (blue) and 2 K (black), along the in-plane (squares data points) and out-of-plane (circle data points) orientations. The solid lines represents the Kittel fitting for both orientations giving straight line fit along the c-axis (fitted with Eq. \ref{easy_axis})  and square-root dependence along ab-plane (fitted with Eq. \ref{hard_axis})  i.e. easy and hard axis, respectively. The dotted line along in-plane orientation show an expected spectrum of frequency dependence for a single domain case.}
\label{fig:Kittel_relation_inset}  
\end{figure}

The broadband FMR, for externally applied magnetic fields along the in-plane and out-of-plane directions for 60 K, 30 K, and 2 K  are shown in Fig. \ref{fig:Kittel_relation_inset}. The inset shows an exemplary experimental resonance spectra. The resonance is described by the derivation of the dynamic susceptibility from the Landau-Lifshitz and Gilbert equation, and hence it is analyzed by fitting a linear combination of symmetric and anti-symmetric Lorentzian functions in order to determine the resonance field, $H_{\textrm{r}}$ (i.e. peak position)\cite{celinski1997using}. The frequency dependence of $H_{\textrm{r}}$ exhibits the typical characteristic of magnetisation dynamics in a ferromagnet with easy and hard magnetic axis. The easy axis in this case is along the c-axis (out-of-plane), which shows a linear frequency dependence with a non-zero y-intercept corresponding to the anisotropy field. The ab-plane (in-plane) resonance spectra shows a non-linear frequency dependence with the x-intercept increasing in value as the temperature is decreased from 60 K to 2 K, exhibiting an increase in the uniaxial magnetocrystalline anisotropy. The dotted lines show the expected resonance spectra of a uniform ferromagnetic resonance in single domain crystals. An important observation in the extracted FMR Kittel relations, is the difference in slope for in-plane and out-of-plane orientations. Additionally, the slope for the two orientations shows a temperature dependence. This suggests that the spectroscopic g-factor is not isotropic for CGT (discussed in more detail below).

In the following we discuss the fitting equations derived by the Smit-Beljers-Suhl approach\cite{smit1955ferromagnetic, suhl1955ferromagnetic}, which are used to fit the experimental data along the in-plane and out-of-plane directions in Fig. \ref{fig:Kittel_relation_inset}. The spatial distribution of the free energy density, \textit{F}, is probed by the FMR experiment. The free energy density for the CGT system excluding exchange energy (in-plane dimensions of 2 x 1.5 mm and thickness of 0.3 mm, i.e. in the out-of-plane direction) reads

\begin{equation}
F = -\textbf{M}\cdot\textbf{H} + 2\pi(\textbf{M} \cdot \textbf{n})^2 - K_{\textrm{u}}(\textbf{M}\hspace{0.05cm}\cdot\hspace{0.05cm} \frac{\textbf{z}}{M_{\textrm{s}}})^2,
\label{free_energy}
\end{equation} 

where $\textbf{H}$ and $\textbf{M}$ are the external magnetic field and magnetisation vectors, respectively. The first term in Eq. \ref{free_energy} is the Zeeman energy, the second term is the demagnetisation energy and the third term is the uniaxial magnetic anisotropy term, with $K_{\textrm{u}}$, the magnetocrystalline anisotropy constant, $M_{\textrm{s}}$, the saturation magnetisation, and \textbf{n} and \textbf{z} are the unit vectors normal to the surface of the sample and orientated along the easy axis (c-axis), respectively. \textbf{H} and \textbf{M} are then written as 

\begin{equation}
\textbf{H}= H 
\begin{pmatrix}
\sin(\theta_{\textrm{H}})\cos(\phi_{\textrm{H}}) \\
\sin(\theta_{\textrm{H}})\sin(\phi_{\textrm{H}}) \\
\cos(\theta_{\textrm{H}})
\end{pmatrix}\hspace{0.2cm},
\label{field_vector}
\end{equation}

\begin{equation}
\textbf{M} = M_{\textrm{s}}
\begin{pmatrix}
\sin(\theta_{\textrm{M}})\cos(\phi_{\textrm{M}}) \\
\sin(\theta_{\textrm{M}})\sin(\phi_{\textrm{M}}) \\
\cos(\theta_{\textrm{M}})
\end{pmatrix}\hspace{0.2cm},
\label{magnetisation_vector}
\end{equation} 

where $\theta_{\textrm{H/M}}$ and $\phi_{\textrm{H/M}}$ are the polar (accounted from normal to the rectangular sample platelet ab-plane)  and azimuthal (in-plane) angles, respectively. The double derivatives approach of the magnetic free energy with respect to the polar and azimuthal angles (i.e. substituting equations \ref{free_energy}, \ref{field_vector} and \ref{magnetisation_vector} into Eq. \ref{Smit_Beilgers}) is then used to analytically calculate the resonance frequency\cite{smit1955ferromagnetic, suhl1955ferromagnetic}, $\omega = 2\pi f$, as

\begin{equation}
\bigg(\frac{\omega}{\gamma}\bigg)^2 = \frac{1}{M_{\textrm{s}}\sin(\theta_{\textrm{M}})} \bigg [ \frac{\partial^2F}{\partial\theta^2_{\textrm{M}}}\frac{\partial^2F}{\partial\phi^2_{\textrm{M}}} - \bigg(\frac{\partial^2F}{\partial\theta_{\textrm{M}}\partial\phi_{\textrm{M}}}\bigg)\bigg],
\label{Smit_Beilgers}
\end{equation}

where $\gamma$, is the gyromagnetic ratio. The azimuthal angles in the CGT system (i.e. in equation \ref{field_vector} and \ref{magnetisation_vector}) can be set to $\phi_{\textrm{H/M}} = 0$, as the in-plane magnetocrystalline anisotropy energy is negligible. Solving for the perpendicular anisotropy case ($\textbf{n} = \textbf{z}$, along the easy axis i.e. parallel to out of plane c-axis), one arrives at the general Kittel relation for arbitrary polar angles of the external magnetic field with respect to the easy axis of the magnetisation, which reads 

\begin{equation}
\begin{multlined}
\bigg(\frac{\omega}{\gamma}\bigg)^2 = [H_{\textrm{r}} \cos(\theta_{\textrm{H}} - \theta_{\textrm{M}}) - 4\pi M_{\textrm{eff}} \cos^2(\theta_{\textrm{M}})] \\ \times [H_{\textrm{r}} \cos(\theta_{\textrm{H}} - \theta_{\textrm{M}}) - 4\pi M_{\textrm{eff}} \cos(2\theta_{\textrm{M}})] 
\end{multlined},
\end{equation}

where $H_{\textrm{r}}$ is the ferromagnetic resonance field, $4\pi M_{\textrm{eff}} = 4\pi M_{\textrm{s}} - H_{\textrm{u}}$ is the effective demagnetisation field and $H_{\textrm{u}} = 2K_{\textrm{u}}/\mu_0 M_{\textrm{s}}$, the perpendicular anisotropy field, and since the hexagonal CGT lacks the \textit{C}$_4$ symmetry in the unit cell, the fourth order anisotropy constant is not considered in the fitting of the FMR spectra. On resonance, the values for $\theta_{\textrm{M}}$ can be determined by the condition $\partial F / \partial\theta_{\textrm{M}} = 0$ and is given by 

\begin{equation}
\sin(2\theta_{\textrm{M}}) = \frac{2 H_{\textrm{r}}}{4\pi M_{\textrm{s}} - H_{\textrm{u}}}{\sin(\theta_{\textrm{M}} - \theta_{\textrm{H}})}\hspace{0.2cm}.
\label{theta_m_equilibrium}
\end{equation}

The Kittel equations for the in-plane ($\Vert$) and perpendicular ($\perp$) orientations can be determined by setting $\theta_{\textrm{M}}=\theta_{\textrm{H}}$ in a saturated state, and are given by

\begin{equation}
\omega_{(\Vert)} = \gamma \sqrt{H_{\textrm{r}} (H_{\textrm{r}}  + 4\pi M_{\textrm{eff}})} \hspace{0.2cm}, 
\label{hard_axis}
\end{equation} 

\begin{equation}
\omega_{(\perp)} = \gamma (H_{\textrm{r}} - 4\pi M_{\textrm{eff}}) \hspace{0.2cm}.
\label{easy_axis}
\end{equation}

The above equations are used to fit the in-plane and out-of-plane FMR data in Fig. \ref{fig:Kittel_relation_inset},  with spectroscopic g-factor (contained in $\gamma = g \mu_{\textrm{B}}/\hbar$) and $M_{\textrm{eff}}$ as fitting parameters. The extracted parameters from the Kittel relations are also used to calculate the dependence of the resonance frequency as a function of externally applied magnetic field, and the angle-dependent resonance spectrum together with the calculated curves is shown in the supplementary material \footnote{See Supplemental Materials at [URL will be inserted by
publisher] for the angle-dependent FMR results and theoretical DFT magnetic anisotropy energy calculations.}. 

\begin{figure}[h!]
\centering
\subfigure[]{
\includegraphics[width=1\linewidth]{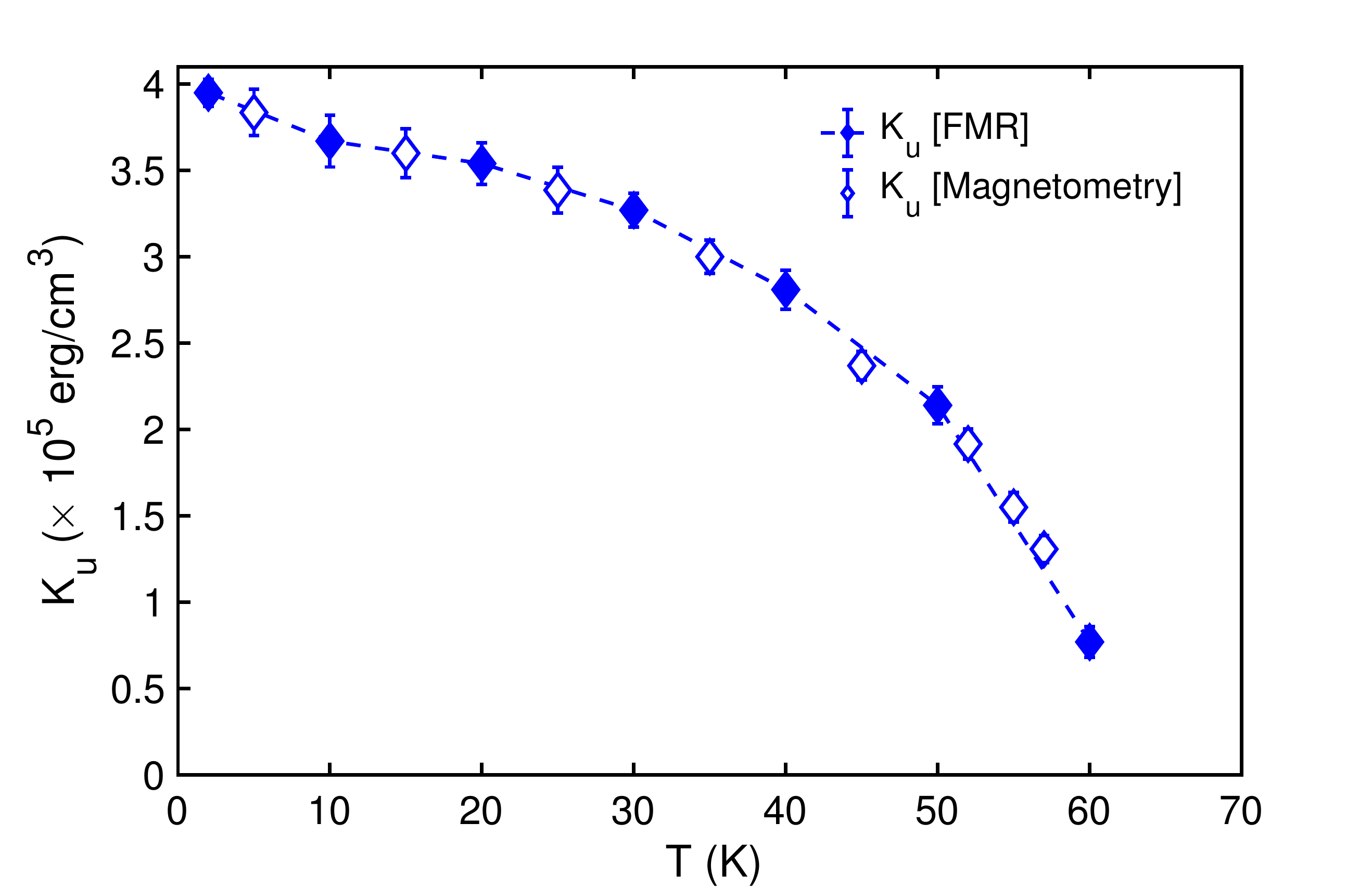}}
\subfigure[]{
\includegraphics[width=1\linewidth]{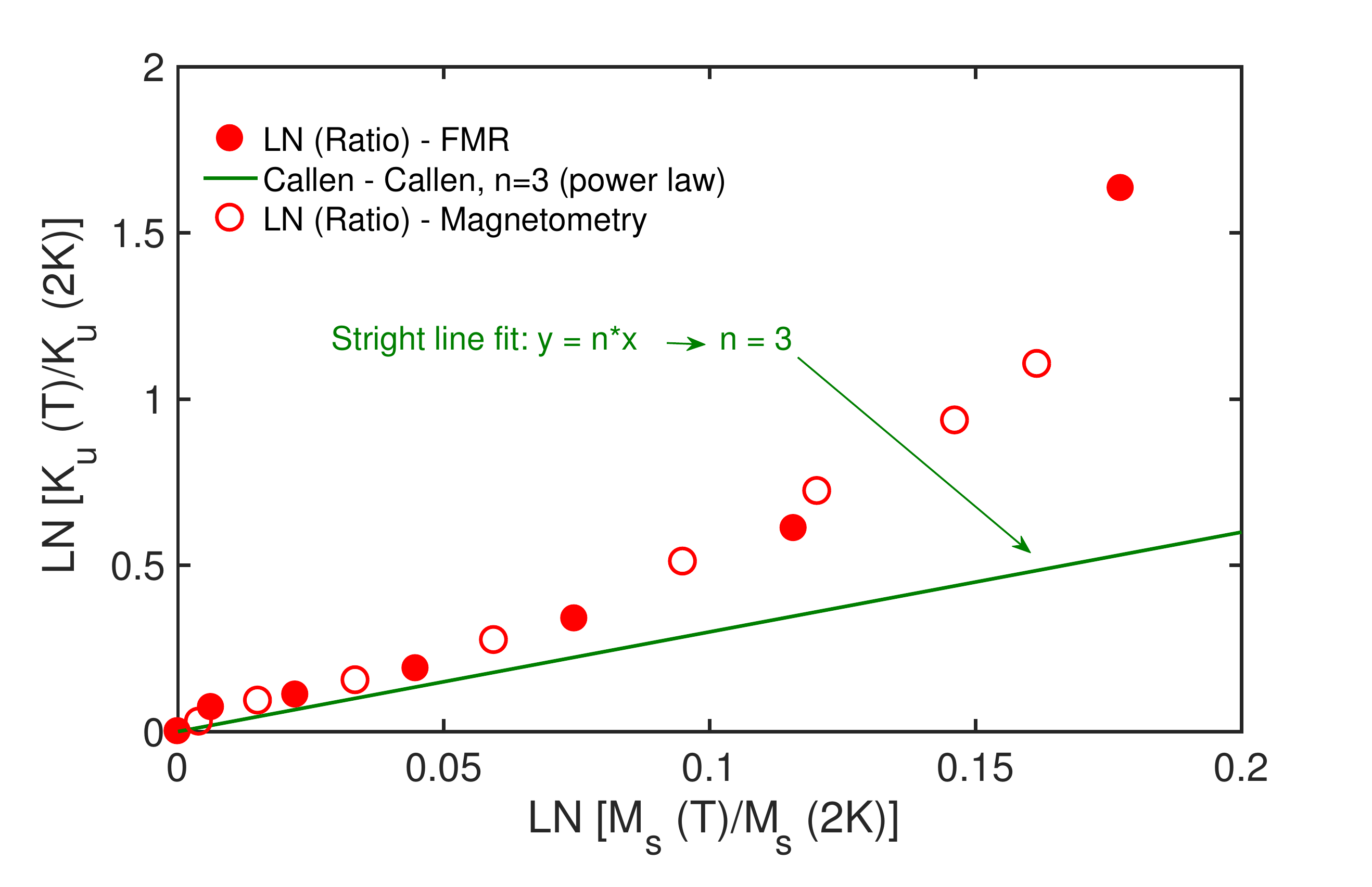}}
\caption{\label{magnetic_anisotropy_energy_graph}(a) Temperature dependence of uniaxial magnetocrystalline anisotropy constant, $K_u$, where the filled points are extracted from the FMR data as discussed in the main text and the non-filled points are extracted from the magnetometry measurements \cite{yang2000uniaxial}. (b) The ratio $K_u (T) / K_u (\textrm{2 K})$ is compared with the ratio $[M_s (T) / M_s (\textrm{2 K})]$ in the absolute value of the logarithm scale. A solid line with the exponent, n = 3, is plotted to highlight that the experimental data does not follow the Callen-Callen behaviour.}
\end{figure}

The temperature dependence of the out-of-plane uniaxial magnetocryatlline anisotropy constant, $K_{\textrm{u}}$, is shown in Fig. \ref{magnetic_anisotropy_energy_graph}(a). This was determined from the least-square fitting of equations \ref{hard_axis} and \ref{easy_axis} to the frequency-dependent FMR spectra. A positive sign convention is used, where $K_{\textrm{u}} > 0$ favors perpendicular magnetic anisotropy. $K_{\textrm{u}}$ is found to be in the range of $(0.77 - 3.95) \times 10^5$ erg/$\textrm{cm}^3$ from 2 K to 60 K, which confirms that the magnetic easy axis in CGT is along the c-axis as seen by the magnetometry data and previous experimental observations \cite{zhang2016magnetic,lin2018pressure}. The reported magnetocrystalline anisotropy energies for other similar layered 2D magnets are $K_{\textrm{u}} = 1.46 \times 10^7$ erg/$\textrm{cm}^3$ at 4 K and $K_{\textrm{u}} = 3.1 \times 10^6$ erg/$\textrm{cm}^3$ at 1.5 K, for Fe$_3$GeTe$_2$ \cite{leon2016magnetic} and CrI$_3$ \cite{dillon1965magnetization}, respectively. CGT is found to be less anisotropic then these systems whereas the ferromagnetic T$_{\textrm{c}}$ of CGT is close to that of CrI$_3$ ($\sim$ 61 K) and Fe$_3$GeTe$_2$ possesses high transition temperature ($\sim$ 223 K). At 2 K, the anisotropy energy value corresponds to 0.24 meV per unit cell of CGT, which is six times smaller than the value, 1.4 meV per unit cell, obtained from the DFT calculations\cite{Note1}. The difference in the experimental and calculated magnetic anisotropy energy can arise as (i) the DFT calculation is done at 0 K, (ii) the structure model used in the calculation is considered to be a defect-free system (the unit cell considered for DFT calculations can be seen in the supplementary material\cite{Note1}).

In Fig.\ref{magnetic_anisotropy_energy_graph}(b), the reduced anisotropy constant, $K_{\textrm{u}}(T)/K_{\textrm{u}}(\textrm{2 K})$, and the reduced magnetisation, $M_{\textrm{s}}(T)/M_{\textrm{s}}(\textrm{2 K})$, at different temperatures are shown for the CGT single crystal. The theory of the temperature dependence of magnetic anisotropies was developed by Callen and Callen, and others\cite{van1937anisotropy,zener1954classical,callen1966present}. It predicts that the magnetocrystalline anisotropy constant as a function of temperature for uniaxial and cubic systems distinctly differs depending on the crystal symmetry and on the degree of correlations between the direction of neighbouring spins. The famous Callen-Callen power law based on the single-ion anisotropy model relates the temperature dependence of $K_{\textrm{u}}$ to $M_{\textrm{s}}$ in the low temperature limit ($\textrm{T} << \textrm{T}_{\textrm{c}}$) as 

\begin{equation}
\frac{K_{\textrm{u}} (T)}{K_{\textrm{u}} (0)} = \bigg[\frac{M_{\textrm{s}} (T)}{M_{\textrm{s}} (0)}\bigg]^{n}\hspace{0.2cm},
\label{callen_callen}
\end{equation}

where the exponent, $n = l(l+ 1)/2 $, depends on the crystal symmetry and degree of correlations between adjacent spins, and \textit{l} being the order of spherical harmonics and it describes the angular dependence of the local anisotropy. In the case of uniaxial anisotropy, the exponent is predicted to be $n = 3$ and it suggests a single-ion origin of magnetic anisotropy. 

An unexpected behaviour of the scaling between the temperature dependence of the anisotropy constant and magnetisation is found in hexagonal CGT. This can be seen in Fig. \ref{magnetic_anisotropy_energy_graph}(b), where the experimental data departs from the trend of a straight line (plotted in logarithmic scale) with a slope of $ n = 3$, predicted by the Callen-Callen power law for uniaxial systems. In simple ferromagnetic systems such as ultrathin Fe films\cite{zakeri2006power}, the Callen-Callen power law for uniaxial anisotropy has been experimentally verified, across the whole temperature range below T$_{\textrm{c}}$, with an exponent $n = 2.9$, and it is considered as a good model for ferromagnets with localised moments. In the past, deviations from the expected scaling exponent of $n = 3$ have been reported in complex systems containing a non-magnetic material with a large spin-orbit coupling, which can contribute to magnetic anisotropy without having a significant effect on other magnetic properties \cite{skomski2003finite, mryasov2005temperature,truong2014evidence}. In highly anisotropic barium ferrite systems, $K_{\textrm{u}}$ was found to be linearly proportional to $M_{\textrm{s}}$, clearly inconsistent from the theoretical predictions \cite{wang2011unusual}. The departure from the Callen-Callen theory for these reported systems is suggested due to the violation of simple assumptions considered in the theory\cite{chatterjee2014temperature}, (i) magnetisation origin in the material coming from single-ions with localised magnetic moments, (ii) spin-orbit coupling being regarded as a small perturbation to the exchange coupling and (iii) temperature dependence of anisotropy and magnetisation having the same origin. Recent first principles calculations in CGT predict the interplay between single-ion anisotropy and Kitaev interaction \cite{xu2018interplay}. They are found to be induced by the off-site spin-orbit coupling of the heavy ligand, i.e. Te atoms. Another recent study \cite{kim2019giant}, investigating the origin of magnetic anisotropy in layered ferromagnetic Cr compounds, suggests that an additional magnetic exchange anisotropy induced by the spin-orbit coupling from the ligand p-orbital through superexchange mechanism plays a crucial role in determining the magnetocrystalline anisotropy in CGT and other layered systems such as CrI$_3$ and CrSiTe$_3$. Hence, CGT is a complex magnetic system with no simple correlation between the thermal behaviour of anisotropy and magnetisation.

\begin{figure}[h]
\centering
\includegraphics[width=1.0\linewidth]{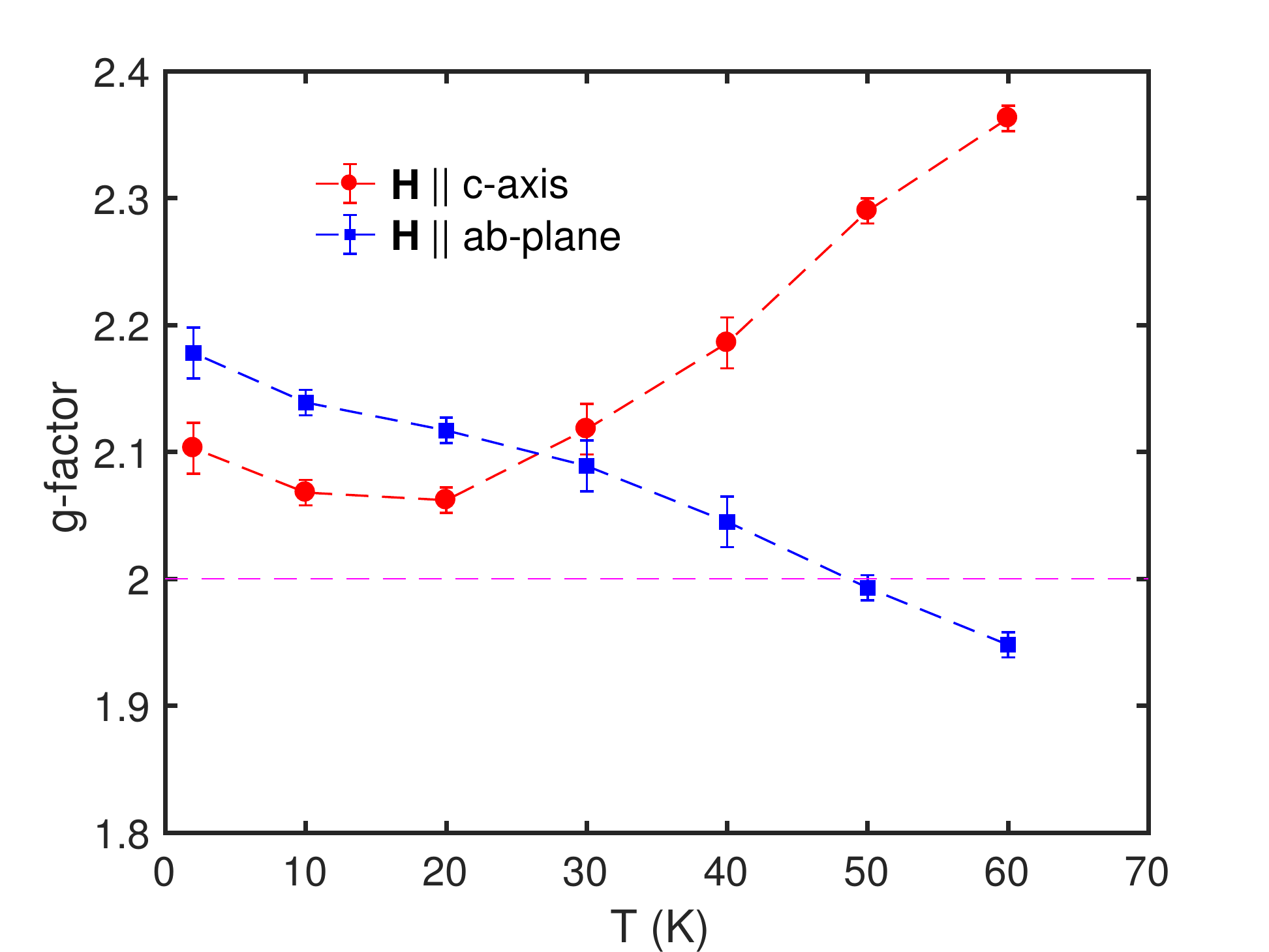}
\caption{Temperature dependence of the spectroscopic g-factor with external magnetic field along the c-axis (easy-axis, red circles) and the ab-plane (hard-axis, blue squares). The dashed horizontal line (magenta) at $g = 2$ indicates a case of pure spin magnetism where the orbital moment is completely quenched.} 
\label{g_factor} 
\end{figure}

The temperature dependence of the spectroscopic g-factor for the in-plane (ab-plane) and out-of-plane (c-axis) orientations,  can be seen in Fig. \ref{g_factor}. There is contrasting behaviour in the temperature dependence of g-factor, where \textit{g} increases with decreasing temperature in the in-plane (hard-axis) direction, and it decreases in the out-of-plane (easy-axis) direction as the temperature decreases. At 2 K, $g_{\Vert} = 2.18 \pm 0.02$ (ab-plane) and $g_{\perp} = 2.10 \pm 0.01$ (c-axis), indicating that the g-factor is anisotropic in CGT and shows a crossing between 20 K and 30 K. The g-factor is determined by extracting the proportionality of the gyromagnetic ratio in Eqs. (\ref{hard_axis}) and (\ref{easy_axis}), given by $\gamma=g \mu_{\textrm{B}} / \hbar$, where $\mu_{\textrm{B}}$ is the Bohr magneton and $\hbar$ is the reduced Planck\textsc{\char13}s constant. Determining a value of \textit{g} is important since one can find the relative spin and orbital moments of a material by using the well-known relation\cite{kittel1949gyromagnetic}

\begin{equation}
\frac{\mu_{\textrm{L}}}{\mu_{\textrm{S}}} = \frac{g -2}{2}\hspace{0.2cm},
\end{equation}

where $\mu_{\textrm{L}}$ is the orbital moment per spin and $\mu_{\textrm{S}} = \mu_{\textrm{B}}$ is the spin moment. The deviation of the spectrocopic factor from the electron\textsc{\char13}s g-factor of $g = 2$ suggests an orbital contribution to the magnetisation\cite{shaw2013precise}. The \textit{g} value of the free electron case is considered in systems where the orbital moment is assumed to be nearly completely quenched due to the high symmetry of the bulk crystals, and hence magnetism in such systems is described in terms of pure spin magnetism. The non-spherical charge distribution in the d-shells can prevent the complete quenching of the orbital momentum\cite{baberschke2001anisotropy}, i.e. $L_{\textrm{z}} \neq 0$. An orbital moment of 8 \%  and 5 \% for $\mu_{\textrm{L}} / \mu_{\textrm{S}}$ along the in-plane and out-of-plane orientations, respectively, is found in CGT at 2 K. The small orbital moment anisotropy could be explained by the presence of the spin-orbit interaction from the d-orbitals of the Cr atoms as well as the off-site spin-orbit coupling of p-orbitals of Te atoms\cite{Note1}, giving \textit{g} a tensor character i.e. dependence on the crystallographic directions. The g-factor extraction from the FMR experiment here only gives a relative insight into the orbital moment contribution to the magnetism in CGT. One can determine the orbital moment, $\mu_{\textrm{L}}$, and its anisotropy from the x-ray magnetic circular dichroism (XMCD) experiments at synchrotron facilities in order to further improve the understanding of orbital magnetism in these layered van der Waals magnetic systems\cite{thole1985strong, altarelli1993orbital,frisk2018magnetic}. In fact, the study referenced previously \cite{kim2019giant} looking into the origin of magnetic anisotropy in layered 2D systems also carries out an initial XMCD investigation of Cr $L_{\textrm{2,3}}$-edge in CGT at 20 K. They report a sizable anisotropy in the orbital angular momentum (i.e. related to the orbital moment), where $L_{\textrm{c}} = - 0.045$, $L_{\textrm{ab}} = - 0.052$ and $\Delta L = 0.007$. This work complements our results of the g-factor anisotropy obtained in the FMR experiment, where a deviation from $g = 2$ confirms presence of orbital angular momentum in CGT. 

\begin{figure}[htb!]
\centering
\includegraphics[]{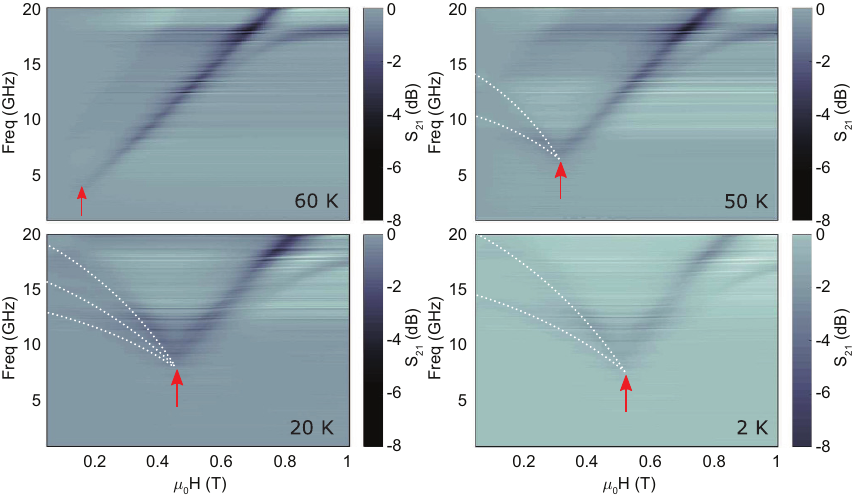}
\caption{Temperature evolution of the frequency-dependent ferromagnetic resonance spectra (in-plane orientation) from 60 K to 2 K. The red arrow indicates the saturation field point in the system where the system goes from multi-domain to single domain state. The dotted white lines do not represent any fittings and are used only for guidance.}
\label{multidomain_picture}   
\end{figure}

In Fig. \ref{multidomain_picture}, the temperature evolution from 60 K to 2 K of the frequency-dependent FMR spectrum in the in-plane (hard-axis) orientation is shown. The red arrows indicate the saturation field points for the CGT system, where a transition from the multi-domain magnetic states to a single domain state occurs in the sample\cite{luhrmann1993high}. At 60K, there is no sign of domain-mode resonance but as the temperature decreases, the strength of the perpendicular uniaxial magnetic anisotropy increases, and the multiple domain-mode resonances (dotted white lines) appear. Qualitatively, this shows the presence of multi-domain structures in bulk CGT \cite{sigal1977ferromagnetic}. The FMR setup used in this experiment is sensitive to observe the resonance of these domain structures. The minimum feature of the resonance frequency (red arrow) increases with decreasing temperature. This type of feature is markedly different for samples with a single magnetic domain state\cite{zakeri2006magnetic}, where the frequency-dependent resonance spectra has a x-intercept along the hard axis of magnetisation.

The \textit{M-H} curves in Fig (1a \& 1b) show no hysteresis behaviour, but this does not reflect signatures of a single domain system below the saturation state for CGT. The remanent magnetisation at zero external magnetic field does not have any net magnetisation, indicating magnetic moment cancellation due to multi-domain states with up-and-down flux closures. This seems to be common for layered van der Waals magnets i.e. exhibiting a soft magnetic behaviour of a linear \textit{M-H} curve before saturation \cite{mcguire2017magnetic, mcguire2015coupling}. The experimental evidence of the presence of multiple magnetic domain structures in bulk CGT along the in-plane sample orientation has been confirmed recently using magnetic force microscopy\cite{guo2018multiple}. The observed domain structures were classified with different symmetry types and showed an evolution from one symmetry type to another as function of both temperature and external magnetic field. This reaffirms that the features seen in the magnetisation dynamics experiment below the saturation field (Fig. \ref{multidomain_picture}) corresponds to dynamics of multiple magnetic domain structures. Although, the presence of multi-domains are realised by the FMR experiment, it is not possible to obtain the symmetry and type of structures for these domains by a resonance experiment alone. Hence, one has to be careful in extracting any magnetic parameters in the bulk layered CGT system at low magnetic fields as a small dynamic demagnetising field can exist from the interaction of different magnetic domains with each other and also with that of domain wall boundaries. 

\section{Conclusion}

In summary, we studied the FMR behaviour in bulk two dimensional CGT. In particular, broadband FMR experiments were performed within the temperature range of 60 K - 2 K with external magnetic fields along the in-plane and out-of-plane orientations. Additionally, the angular dependence of the FMR spectra was examined. We focused on the temperature dependence of the magnetic anisotropy energy, spectroscopic g-factor and the multi-domain resonance features (observed below the saturation fields along the ab-plane). A uniaxial perpendicular magnetic anisotropy is found with easy axis parallel to the c-axis (out-of-plane). The scaling of the normalised magnetocrystalline anisotropy constant and saturation magnetisation showed a deviation from the theoretically predicted Callen-Callen power law. The presence of spin orbit coupling from d-orbitals of Cr atoms and an off-site spin orbit coupling from p-orbitals of the ligand Te atoms can be a contributing factor to this contradicting behaviour of power law dependence of magnetic anisotropy to magnetisation. The obtained g-factor showed an anisotropic response, i.e.  a cystallographic dependence, and at 2 K it was found to be $g_{\Vert} = 2.18 \pm 0.02$ (ab-plane) and $g_{\perp} = 2.10 \pm 0.01$ (c-axis). The determined g-factor values are greater than 2, again, indicating that the orbital magnetism plays an important role in bulk CGT when considering magnetic interactions in this system. A small anisotropy in the orbital magnetic moment relative to the spin moment is found, and a detailed x-ray magnetic circular dichroism study could give further insight into the orbital moment anisotropy. Finally, the presence of multi-domain structures is qualitatively confirmed as domain-mode resonance phenomena are observed along the ab-plane (in-plane) orientation.
 
\section*{Acknowledgements}

This research was supported by the Leverhulme Trust (grant No. RPG-2016-391), the Engineering and Physical Sciences Research Council (EPSRC) through UNDEDD (EP/K025945/1) as well as by the European Union's Horizon 2020 research and innovation programme under Grant Agreement Nos. 688539 (MOS-QUITO) and 771493 (LOQO-MOTIONS). We would like to thank Prof. M. Farle and Dr. J. Baker for the helpful discussions.

%


\clearpage

\widetext
\begin{center}
\textbf{\large Supplementary Material: A spin dynamics study in layered van der Waals single crystal, Cr$_2$Ge$_2$Te$_6$}
\end{center}

\setcounter{equation}{0}
\setcounter{figure}{0}
\setcounter{table}{0}
\setcounter{page}{1}
\makeatletter
\renewcommand{\theequation}{S\arabic{equation}}
\renewcommand{\thefigure}{S\arabic{figure}}
\renewcommand{\bibnumfmt}[1]{[S#1]}
\renewcommand{\citenumfont}[1]{S#1}

\setcounter{equation}{0}
\setcounter{figure}{0}
\setcounter{table}{0}
\setcounter{page}{1}
\makeatletter
\renewcommand{\theequation}{S\arabic{equation}}
\renewcommand{\thefigure}{S\arabic{figure}}
\renewcommand{\bibnumfmt}[1]{[S#1]}
\renewcommand{\citenumfont}[1]{S#1}

\section{Density Functional Theory Calculations}

\subsection{Computational details}
The first-principle calculations are performed using the projector augmented-wave (PAW)\cite{blochl1994projector} method as implemented in the Vienna ab initio simulation package (VASP)\cite{kresse1996g}. The exchange-correction interaction is treated with the Perdew-Burke-Ernzerhof (PBE) form of the generalized gradient approximation(GGA)\cite{perdew1996generalized}. A large plane-wave cut-off energy of 500 eV is used throughout. The van der Waals correction of layered CrGeTe$_3$ is included using the method of Grimme (DFT-D2)\cite{grimme2006semiempirical}. The GGA$+$U method\cite{liechtenstein1995density} is adopted to improve the description of on-site Coulomb interaction between Cr 3d electrons. It was previously reported that the appropriate U value should be within the range of $0.2 < \textrm{U} < 1.7$ eV \cite{gong2017discovery,fang2018large,xu2018interplay}. Here, test calculations were performed using $\textrm{U}_{\textrm{eff}} = 0, 0.5, 1.0, 1.5, 2.0, 2.5$ and $3.0$ eV, respectively, and we choose U $= 1.5$ eV, based on the fact that (i) the calculated lattice constants (a $= 6.901$ and c $= 20.143 \angstrom$) are in good agreement with the experimental data (a $= 6.820$ and c $= 20.371 \angstrom$), as shown in Table 1; (ii) the intralayer magnetic coupling of bulk CrGeTe3 is ferromagnetic, and the ferromagnetic state is more stable due to low magnetic coupling energy, compared to that obtained using other values of $\textrm{U}_{\textrm{eff}}$ (see Table 1). Both lattice parameter and atomic position are optimized under the constraint of the symmetry until the energy and residual force on each atom converge to less than $1 \times 10^{-6}$ eV and $0.01$ eV/$\angstrom$, respectively. The magneto-crystalline anisotropy energy (MAE) is extracted by calculating the energy difference between states with all spins along the x direction [100] and along the z direction [001], respectively, in the bulk with spin-orbit coupling (SOC). We also calculated the total energy of bulk CrGeTe$_3$ along the y direction [010] and found that its energy is identical to that along the x direction. Thus, in the following, the [100] direction is used for the in-plane direction of magnetization. 

\begin{table*}[h!]

\begin{ruledtabular}
\begin{tabular}{ccccccccc} 

\textbf{U$_{\textrm{eff}}$} & \textbf{a} & \textbf{error} & \textbf{c} & \textbf{error} & \textbf{M$_{\textrm{Cr}}$}  & \textbf{M$_{\textrm{tot}}$} &\textbf{band gap} & \textbf{E$_{\textrm{FM}}$ - E$_{\textrm{AFM}}$ }\\ 
(eV)   & ($\angstrom$)& ($\%$) & ($\angstrom$)&($\%$) & ($\mu_{\textrm{B}}$) & ($\mu_{\textrm{B}}$) & (eV)& (eV) \\
\hline
0   & 6.889 & 1.011 & 19.936 & -2.136 & 3.066 & 18.00 & 0.073 & -0.299 \\ 
0.5 & 6.890 & 1.021 & 20.031 & -1.669 & 3.111 & 18.00 & 0.162 & -0.292 \\
1.0 & 6.895 & 1.099 & 20.079 & -1.433 & 3.184 & 18.00 & 0.147 & -0.300 \\
1.5 & 6.901 & 1.192 & 20.143 & -1.119 & 3.257 & 18.00 & 0.152 & -0.303 \\
2.0 & 6.907 & 1.278 & 20.208 & -0.801 & 3.329 & 18.00 & 0.12  & -0.301\\
2.5 & 6.914 & 1.376 & 20.253 & -0.577 & 3.399 & 18.00 & 0.10  & -0.297\\ 
3.0 & 6.922 & 1.494 & 20.284 & -0.429 & 3.466 & 18.00 & 0.068 & -0.289\\ 
\textbf{Exp.}\cite{li2018electronic,sun2018effects} & 6.820 & --- & 20.371 & --- & 3.0 & --- & 0.38 & ---\\

\end{tabular}
\end{ruledtabular}
\caption{Calculated lattice constants, magnetic moment, band gap and intralayer magnetic coupling energy (the energy difference between the ferromagnetic (E$_{\textrm{FM}}$) and anti-ferromagnetic (E$_{\textrm{AFM}}$) configurations) of bulk CrGeTe$_3$ with different U$_{\textrm{eff}}$ values. M$_{\textrm{tot}}$ and M$_{\textrm{Cr}}$ stand for the magnetic moment of total and a Cr atom, respectively. The error refers to the difference between the calculated and experimental lattice constant of bulk CrGeTe$_3$. For comparison, the available experimental values are also listed at the last line of the table.}
\end{table*}

\subsection{Results \& Discussion}
\begin{figure}[htb!]
\centering
\includegraphics[width=0.8\linewidth]{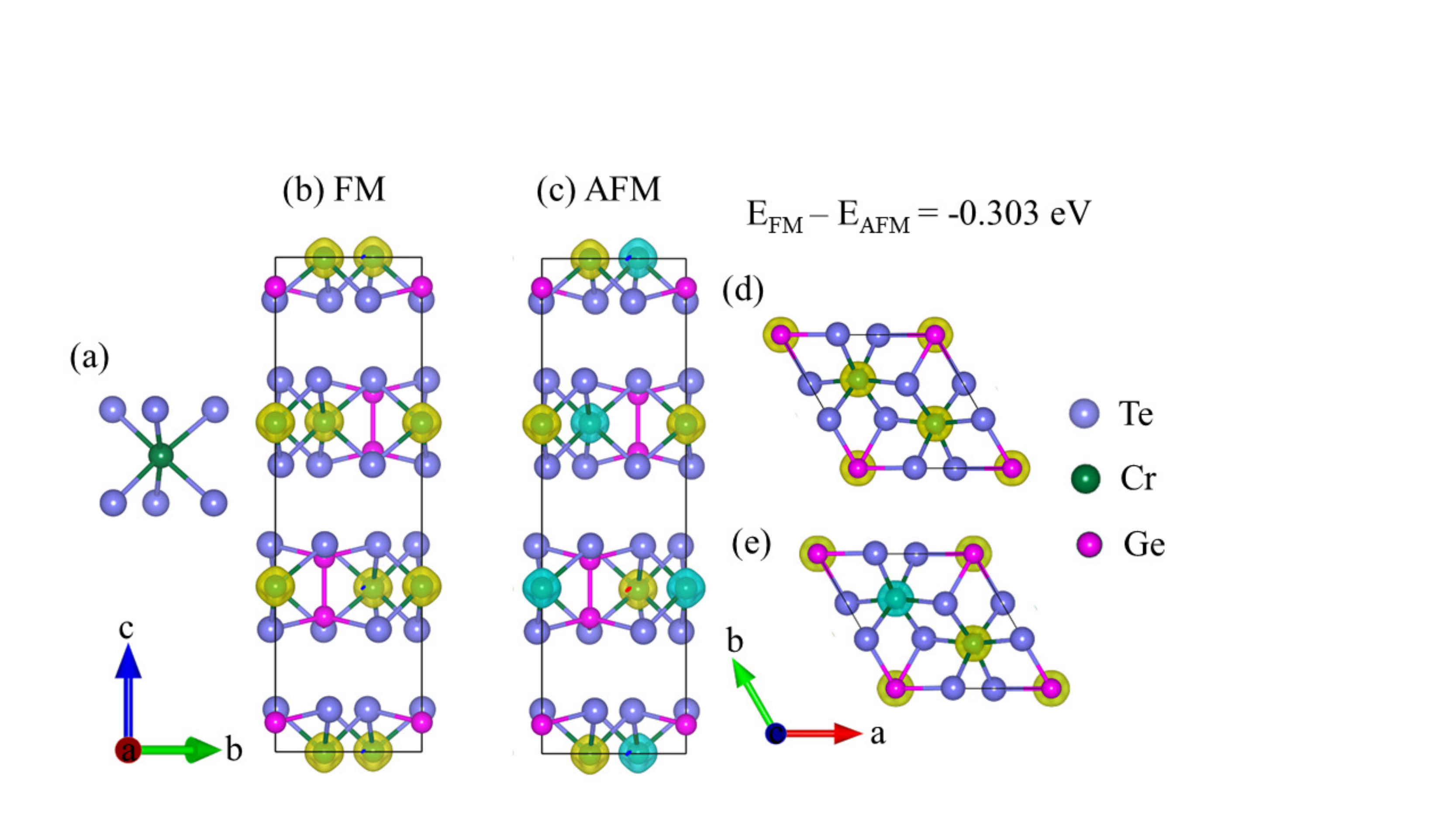}
\caption{(a) Side view of local structure of bulk CrGeTe$_3$, showing that one Cr atom is octahedrally coordinated by six Te atoms. (b) and (d) Side and top view, respectively, of spin polarized charge densities of bulk CrGeTe$_3$ in the FM configuration; (c) and (e), the same information in the AFM configuration. The yellow color and blue color represent the charge density of spin up and down, respectively. The isosurface is 0.058 e$\angstrom^{-3}$.}
\label{figure_1}   
\end{figure}

Bulk CrGeTe$_3$ has a hexagonal crystal structure with the R$\bar{3}$ space group, containing 30 atoms per unit cell. To explore the magnetic ground state, we considered the ferromagnetic (FM) and anti-ferromagnetic (AFM) magnetization configurations as shown in Fig. \ref{figure_1}. The total energy calculations reveal that the FM state is the ground state, the AFM configuration is 0.303 eV per unit cell higher in energy than the FM state. Therefore, we discuss only FM state in the following. It is clear from the spin polarized charge densities of bulk CrGeTe$_3$ (Fig. \ref{figure_1}) that the magnetic moment is contributed mainly by the Cr atoms. The calculated  magnetic moment is about 3.26 $\mu_{\textrm{B}}$/Cr which is in good agreement with the experimental value of about 3.0 $\mu_{\textrm{B}}$ \cite{sun2018effects}. It is expected since there are three unpaired electrons in the Cr$^{3+}$ ionic configuration. A small magnetic moment, -0.12 $\mu_{\textrm{B}}$/Te and 0.034 $\mu_{\textrm{B}}$/Ge, respectively, is also induced on the Te and Ge atoms. The total magnetic moment of CrGeTe$_3$ is 3 $\mu_{\textrm{B}}$/formula unit which is in good agreement with the measured magnetization in bulk CrGeTe$_3$ \cite{sun2018effects}. The ferromagnetic coupling in bulk CrGeTe$_3$ is attributed to the near 90$\degree$ Cr-Te-Cr super-exchange interaction between half-filled Cr-t$_{\textrm{2g}}$ and empty Cr-e$_{\textrm{g}}$ states through the Te-p orbitals, against the direct t$_{\textrm{2g}}$-t$_{\textrm{2g}}$ exchange interactions of Cr in the AFM state. This is also supported by our finding of magnetic moment of Te atoms pointing in opposite direction from the magnetic moment of Cr atoms.

Using the spin-orbit coupling implementation of density functional theory (DFT), we estimate the magneto-crystalline anisotropy energy for ferromagnetic bulk CrGeTe$_3$. Figure \ref{figure_2b} shows the magnetic anisotropy energy surface of bulk CrGeTe$_3$, the energy maxima are observed along the [001] direction, meaning that the c-axis is the easy magnetization direction of bulk CrGeTe$_3$ which agrees with the experimental mensuration. The calculated MAE depends on the choice of U$_{\textrm{eff}}$ as shown in Fig. \ref{figure_2a}, and the MAE is 1.4 meV per unit cell at U$_{\textrm{eff}} = 1.5$ eV. In this work, we did not consider the shape anisotropy caused by dipole-dipole interactions which is expected to be small compared to the magneto-crystalline anisotropy \cite{daalderop1990first}.  

\begin{figure}[h!]
\centering
\subfigure[]{
\includegraphics[width=0.35\linewidth]{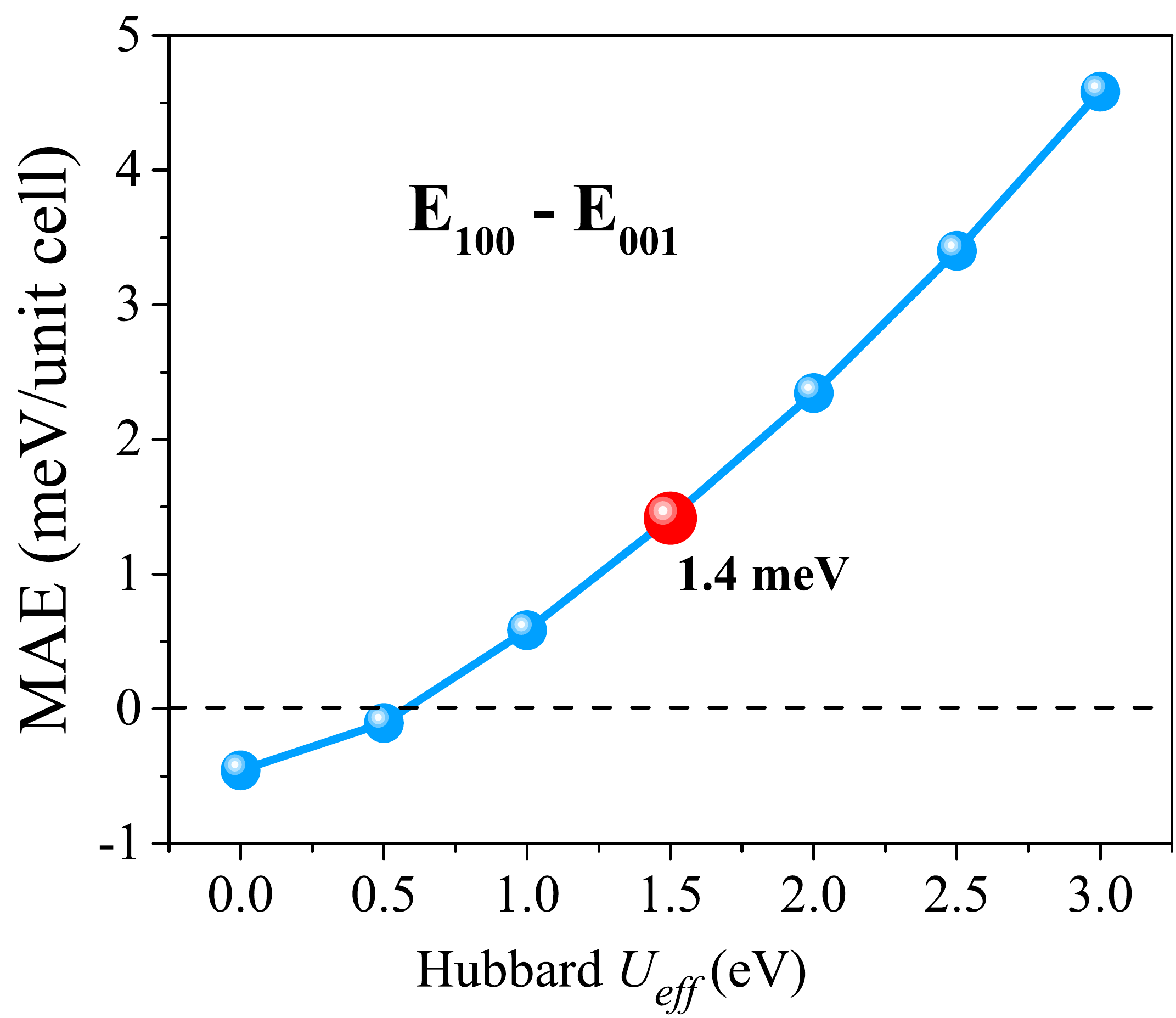}}
\hspace*{1cm}
\subfigure[]{
\includegraphics[width=0.3\linewidth]{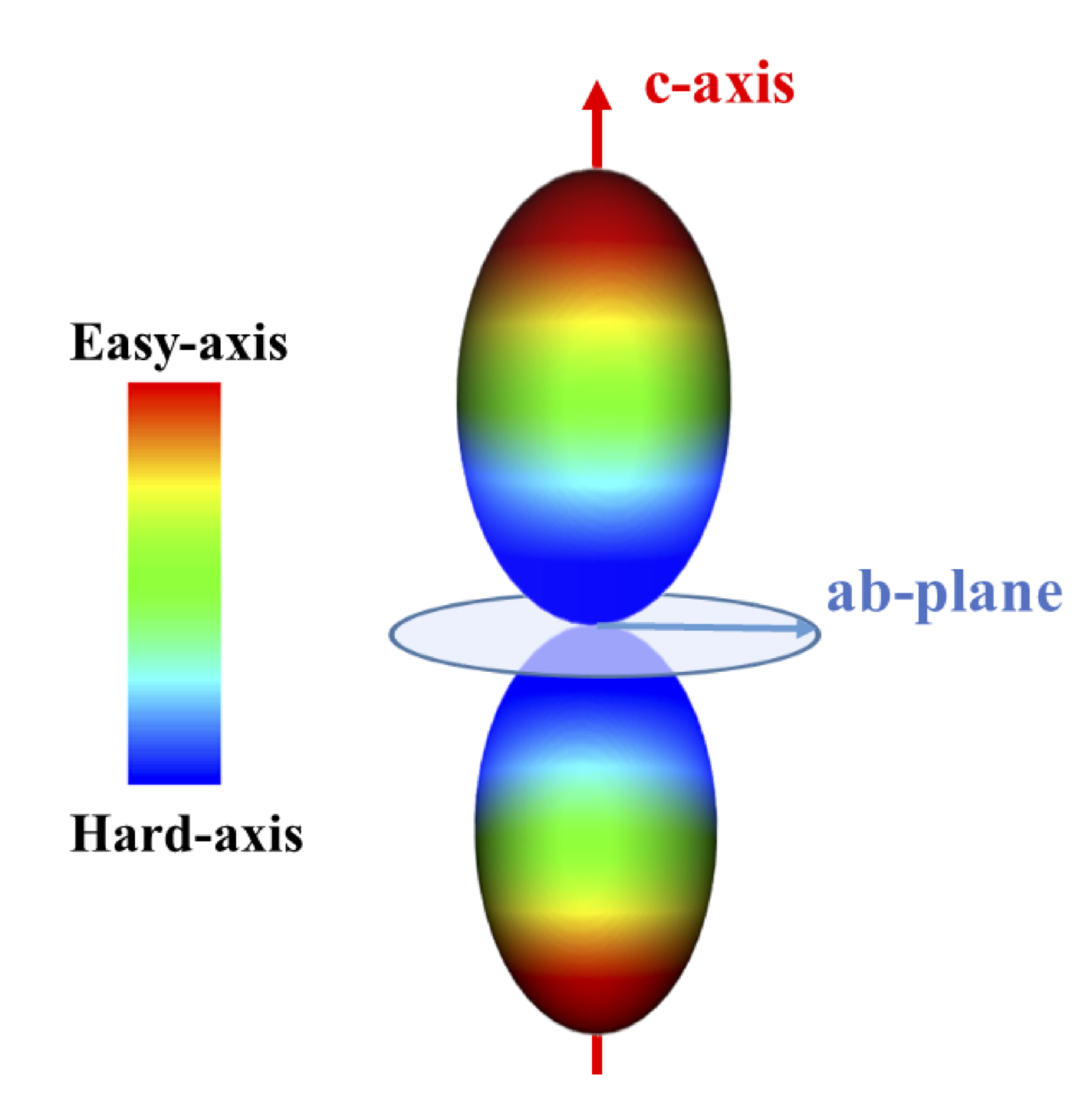}}
\caption{\label{figure_2}(a) Magneto-crystalline anisotropy energy as a function of Hubbard Ueff in bulk of CrGeTe$_3$. Positive MAE value means that the out of plane magnetization is the easy axis. (b) The three-dimensional shape of MAE at U$_{\textrm{eff}} = 1.5$ eV shows that the easy axis of magnetization is the c-axis, that is, [001] direction of the crystallographic cell. The MAE is 1.4 meV which agrees well with the other calculated values of 1.41 meV \cite{fang2018large}.}
\end{figure}

\begin{figure}[h!]
\centering
\subfigure[]{
\includegraphics[width=0.35\linewidth]{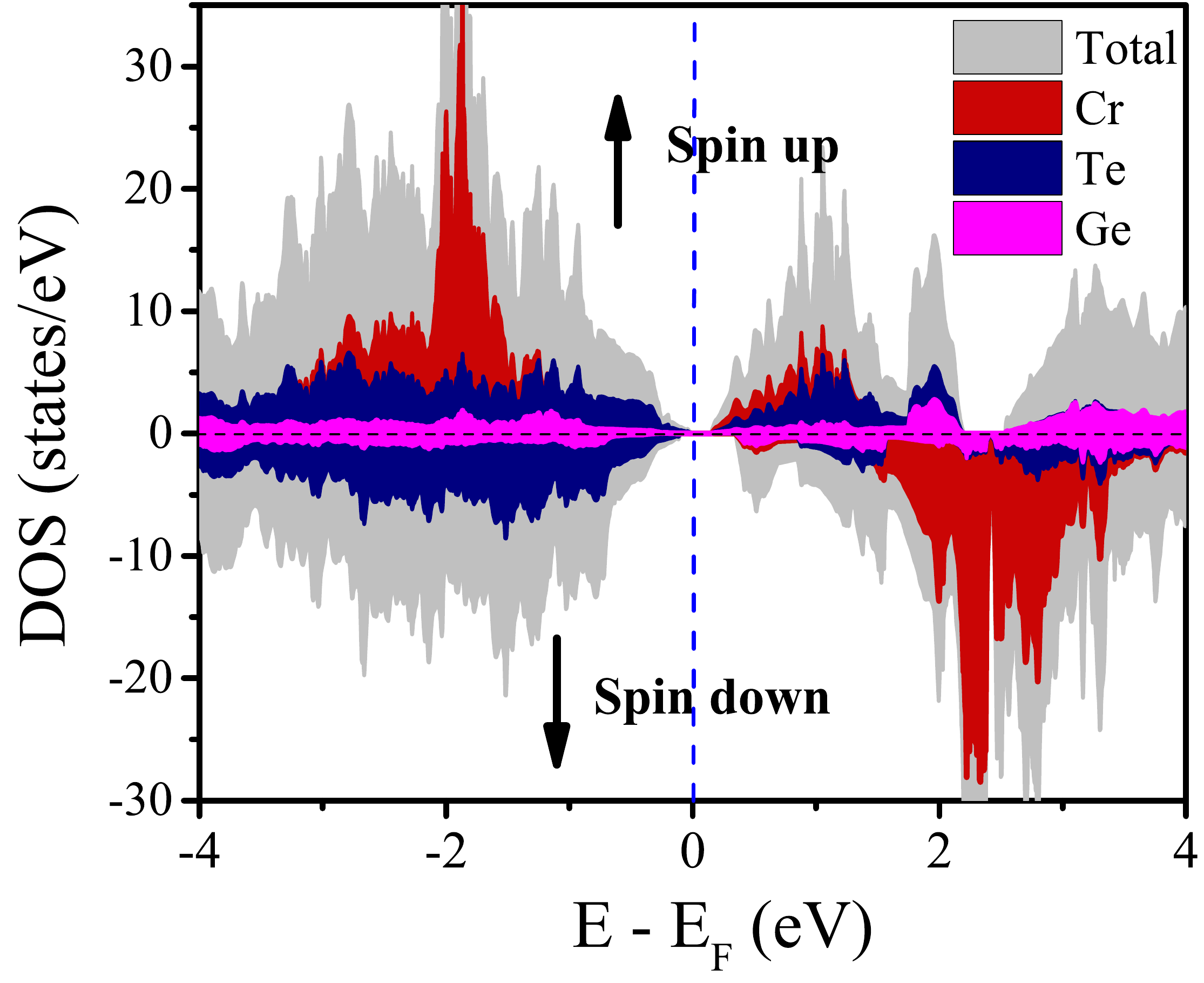}}
\subfigure[]{
\includegraphics[width=0.35\linewidth]{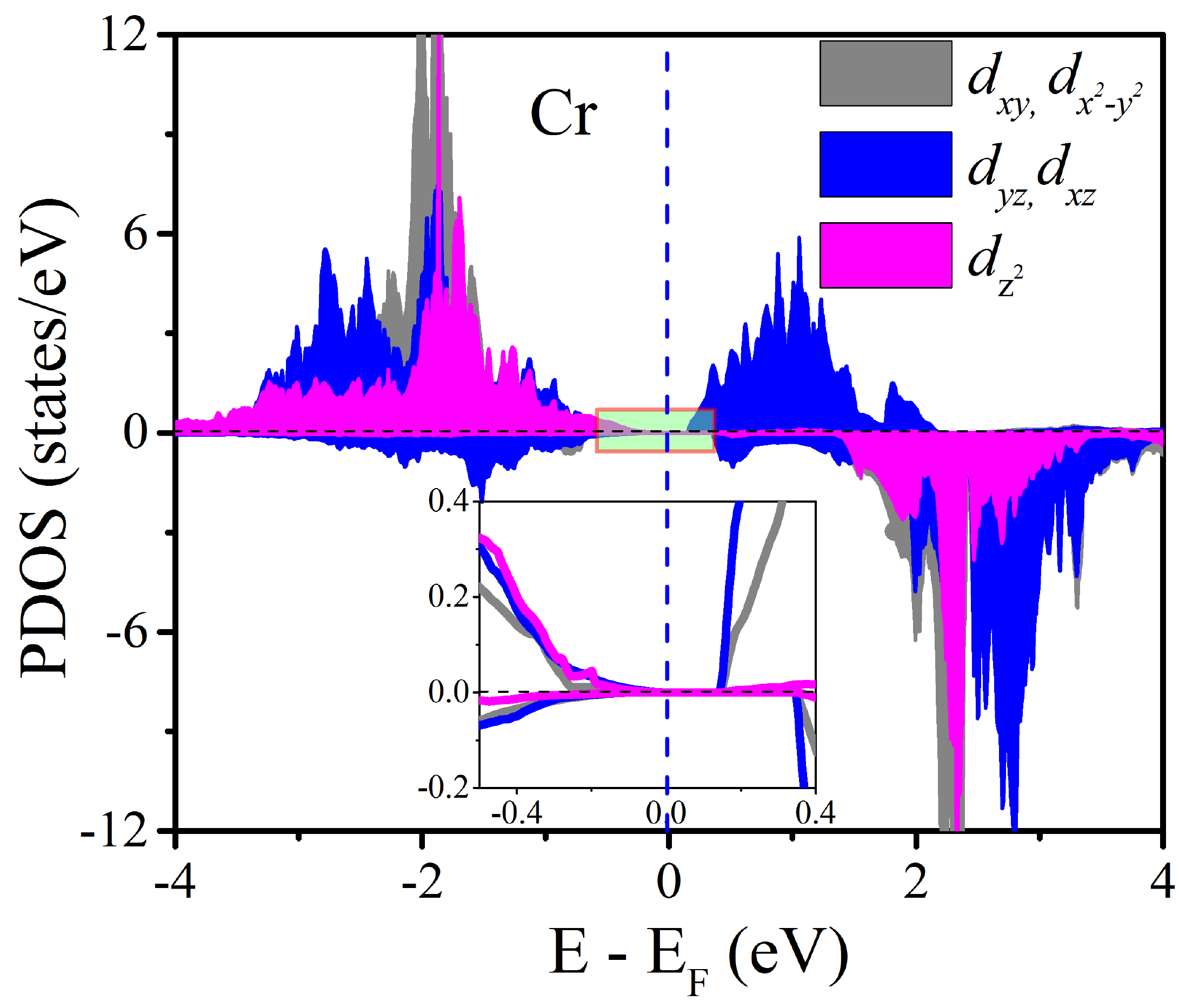}}
\subfigure[]{
\includegraphics[width=0.35\linewidth]{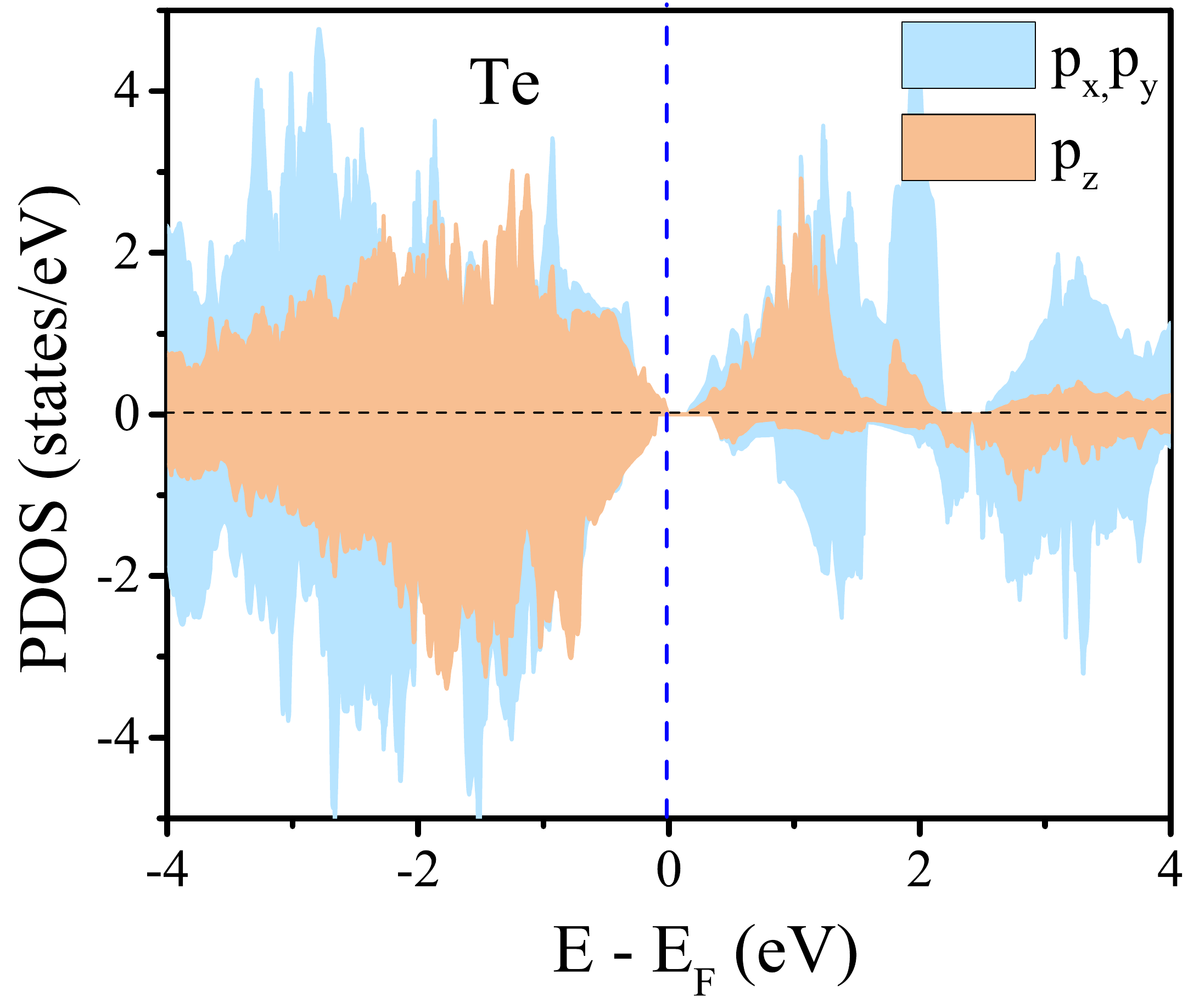}}
\caption{\label{figure_total}GGA+U calculated total DOS of CrGeTe$_3$ (a) and site PDOS of the Cr atoms (b) and the Te atoms (c). The inset in (b) shows an enlarged DOS near the Fermi level. The fermi level is set to zero (denote by blue dashed line).}
\end{figure}

To understand the mechanism of the magnetic anisotropy in bulk CrGeTe$_3$, we calculated its spin polarized total and projected density of state (DOS) and the results are shown in Fig. \ref{figure_total}. It is clear from Fig. \ref{figure_3a} that bulk CrGeTe$_3$ is a ferromagnetic semiconductor with a band gap of $0.15$ eV. The valence band ranging from $-4.0$ to $-0.01$ eV and conduction band ranging from $0.15$ to $4.0$ eV, originated mainly from the Cr d orbitals and Te p orbitals with very small contributions from Ge. Only the PDOS of Cr and Te are shown in Figs. \ref{figure_3b} and \ref{figure_3c}. There are strong orbital hybridizations between Cr d orbitals and Te p orbitals at the valence band maximum (VBM) and conduction band minimum (CBM), which may enhance the magnetic anisotropy due to strong spin-orbit coupling of Te. The energy splitting of the Cr d states around the Fermi level contributes primarily to the crystal field splitting in their octahedrally coordination environment as shown in Fig. \ref{figure_1}(a) (one Cr ion is octahedrally coordinated by six Te atoms). Hence, the band gap in this material is very sensitive to the Hubbard U onsite Cr ions as shown in Table 1. According to the second-order perturbation theory\cite{wang1993first}, the orbitals close to the Fermi energy contribute the most to MAE. Hence, the increase in MAE with the U$_{\textrm{eff}}$ value (see Fig. \ref{figure_2a}) could be due to the enhanced SOC between the d orbitals induced by the decrease of band gap with the increase of U$_{\textrm{eff}}$ value. The majority-spin d$_{\textrm{z}^2}$ states of Cr atom are almost fully occupied. This means that the contribution to the MAE from the SOC between unoccupied majority-spin d$_{\textrm{z}^2}$ states and occupied majority- or minority-spin d states can be neglected and hardly contribute to the MAE. In contrast, the d$_{\textrm{x}^2 - \textrm{y}^2}$, d$_{\textrm{xy}}$, d$_{\textrm{yz}}$ and d$_{\textrm{xz}}$ orbitals of Cr atoms are dominant in both valence band and conduction band, thus their contributions would be large for MAE. Moreover, due to the SOC matrix elements $<\textrm{d}_{\textrm{xz}}|\textrm{H}_{\textrm{SO}}|\textrm{d}_{\textrm{yz}}>$ and  $<$ d$_{\textrm{x}^2 - \textrm{y}^2}|$ H$_{\textrm{SO}}|$ d$_{\textrm{xy}}>$ contribute to the out-of-plane anisotropy,  while $<$ d$_{\textrm{x}^2 - \textrm{y}^2}|$ H$_{\textrm{SO}}|$ d$_{\textrm{yz}}>$, $<$ d$_{\textrm{xy}}|$H$_{\textrm{SO}}|$ d$_{\textrm{xz}}>$, and $<$ d$_{\textrm{z}^2}|$ H$_{\textrm{SO}}|$ d$_{\textrm{yz}}>$ favor an in-plane anisotropy\cite{fang2018large,takayama1976magnetic}, it can be deduced that the SOC interaction from the d$_{\textrm{x}^2 - \textrm{y}^2}$, d$_{\textrm{xy}}$, d$_{\textrm{yz}}$ and d$_{\textrm{xz}}$ orbitals of Cr as well as p orbitals of Te atoms lead to the MAE that prefers the out-of-plane anisotropy in the bulk CrGeTe$_3$ material.

\section{Angular-dependent FMR}

\begin{figure}[htb!]
\centering
\includegraphics[width=0.6\linewidth]{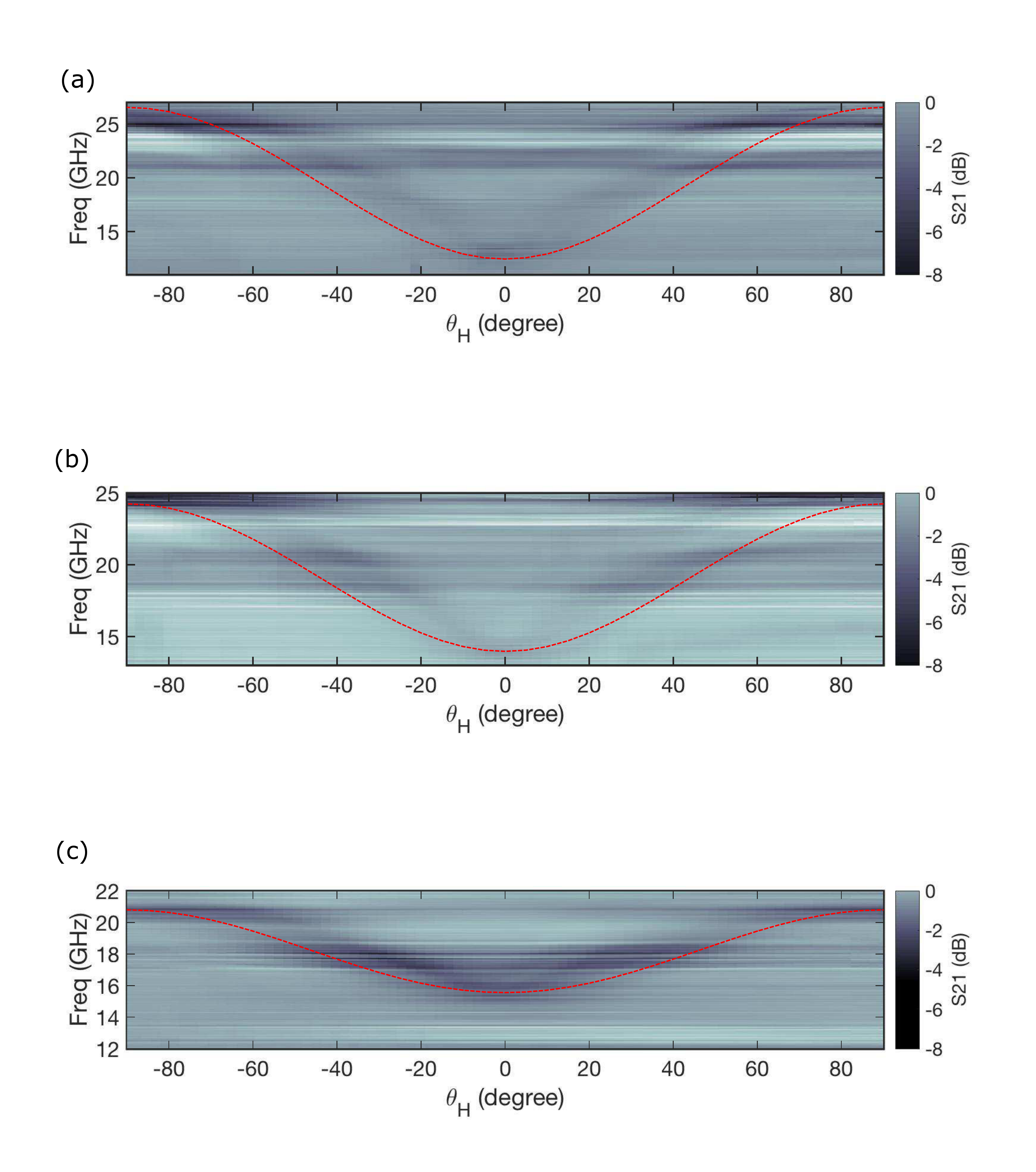}
\caption{Dependence of the ferromagnetic resonance frequency as a function of the applied external magnetic field, $\theta_{\textrm{H}}$, at (a) 2 K , (b) 30K and (c) 60 K. The red line shows the the calculated frequency dependence using the extracted parameters from the Kittel fittings in the main text.}
\label{figure_3}   
\end{figure}

The angular dependence of the resonance frequency with respect to the externally applied magnetic field for 60 K, 30 K and 2 K is shown in Fig. \ref{figure_3}. The red line shows the calculated resonance frequency curve as function of $\theta_{\textrm{H}}$, with ab-plane (hard axis) at $\theta_{\textrm{H}} = 0\degree$ and c-axis (easy axis) at $\theta_{\textrm{H}} = 90\degree$. The resonance frequency is calculated by using equations 1 - 6 in the main text together with the parameters (i.e. $K_u$ and g-factor) extracted from the Kittel relations fitting along the in-plane and out-of-plane orientations. The calculated curves show a good agreement to experimental resonance spectra with a small discrepancy at intermediate angles between in-plane and out-of-plane configurations. This is due to the fact that the spectroscopic g-factor is found to be anisotropic from the Kittel relations and here, it is assumed\cite{stankowski2000anisotropy} as $g^2(\theta_H) = g_{\perp} \cos^2(\theta_H) + g_{\Vert} \sin^2(\theta_H)$, where the $g_{\perp}$ and $g_{\Vert}$ values are given in the main text.

\end{document}